\documentclass[pra,10pt,twocolumn,superscriptaddress,floatfix,showpacs]{revtex4-1}
\usepackage{subfig}
\usepackage{graphicx} 
\usepackage{epstopdf}
\usepackage[utf8]{inputenc}
\usepackage[T1]{fontenc}     
\usepackage[british]{babel}  
\usepackage[sc,osf]{mathpazo}
\usepackage[scaled=0.86]{berasans}  
\usepackage[colorlinks=true, citecolor=blue, urlcolor=blue]{hyperref}  
\usepackage[babel]{microtype}  
\usepackage{amsmath,amssymb,amsthm,bm,amsfonts,mathrsfs,bbm} 

\usepackage{xspace}  
\usepackage{pgfplots}
\usepackage{xcolor,colortbl}
\def\ba{\begin{equation}}
\def\ea{\end{equation}}
\def\bea{\begin{eqnarray}}
\def\eea{\end{eqnarray}}
\def\ben{\begin{equation*}}
\def\een{\end{equation*}}
\def\bean{\begin{eqnarray*}}
\def\eean{\end{eqnarray*}}
\def\bma{\begin{mathletters}}
\def\ema{\end{mathletters}}
\def\bi{\begin{itemize}}
\def\ei{\end{itemize}}
\newcommand{\be}{\begin{equation}}
\newcommand{\ee}{\end{equation}}

\newcommand{\kommentar}[1]{}

\newcommand{\forget}[1]{}

\begin{document}

\title{Any Two Qubit State Has Non Zero Quantum Discord Under Global Unitary Operations}
\author{Kaushiki Mukherjee}
\email{kaushiki_mukherjee@rediffmail.com}
\affiliation{Department of Mathematics, Government Girls' General Degree College, Ekbalpore, Kolkata-700023, India.}
\author{Biswajit Paul}
\email{biswajitpaul4@gmail.com}
\affiliation{Department of Mathematics, Balagarh Bijoykrishna Mahavidyalaya, Balagarh, Dist. - Hooghly-712501, India}
\author{Sumana Karmakar}
\email{sumanakarmakar88@gmail.com}
\affiliation{Department of Mathematics, Heritage Institute of Technology, Anandapur, Kolkata-700107, India.}

\begin{abstract}
Quantum discord is significant in analyzing quantum nonclassicality beyond the paradigm of entanglement. Presently we have explored the effectiveness of global unitary operations in manifesting quantum discord from a general two qubit zero discord state. Apart from the emergence of some obvious concepts such as  absolute classical-quantum, absolute quantum-classical states, more interestingly, it is observed that set of states characterized by absoluteness contains only maximally mixed state. Consequently this marks the peak of effectiveness of global unitary operations in purview of manifesting nonclassicality from arbitrary two qubit state when other standard methods fail to do so. Set of effective global unitaries has been provided in this context. Our observations have direct implications in remote state preparation task.
\end{abstract}

\maketitle

	
\section{Introduction}
Quantum entanglement plays an ubiquitous role in marking deviation of quantum theory from its classical counterpart. Apart from its philosophical importance\cite{horr}, it is one of the most useful nonclassical ingredient from practical viewpoint. Multiple information tasks such as remote entanglement distribution\cite{entd}, secret sharing\cite{ss1,ss2,ss3}, dense coding\cite{dc1,dc2}, teleportation\cite{tele1,tele2}, quantum key generation\cite{key1}, reducing communication complexity\cite{cmp1,cmp2}, etc use entanglement as the resource. However, practically speaking, extreme precise control over quantum measurements together with requirement of isolating the system from environment are essential for generation and use of entanglement. Also, investigation of some quantum information processing protocols using unentangled quantum resources reveal remarkable increase in data processing efficiency\cite{eff1,eff2,eff3,eff4,eff5}. Manifestation of quantum nonclassicality beyond entanglement thus deserves to be analyzed further. Our present study will evolve in this perspective.\\
Pre-existing literature in this direction helps to interpret nonclassical correlations attributable to unentangled quantum systems. In the context of exploiting nonclassical trait beyond quantum entanglement, discord turns out to be one of the most important features of physical systems\cite{modi}. The concept of discord rests upon gap existing between two definitions of mutual information in quantum systems\cite{QD}. Ever since its inception\cite{QD}, quantum discord has been extensively investigated\cite{separable,alter,pati14ii,discordentropy,suffis,pati14iii,suffi,adesso,self}$.\,\textmd{Apart}$ from analysis of its various fundamental aspects, over past few years, study of quantum discord is oriented in exploring its use as a resource in quantum information processing tasks\cite{p1,p2,p3,p4,p5,p6}. \\
In literature there exist various quantum computational models whose gain over corresponding classical analogues rely upon creation of non zero quantum discord. Implementation of some of those protocols such as `deterministic quantum computation with one qubit'(DQC1)\cite{p7,p8,p9,p10} are even found to be classically impossible. Experimental realization of such protocols has also been achieved\cite{eff1,eff2,eff3,eff4}. However, all these studies only indicated utility of this nonclassical feature as a potential resource in quantum information processing(QIP) protocols without analyzing explicitly the dependence of the protocol's efficiency on it. Relation between discord and that of parameter characterizing efficiency of a quantum protocol was studied for some specific QIP protocols such as quantum state merging\cite{p2}, local broadcasting\cite{p3}, dense coding\cite{p2}, remote state preparation\cite{rem1,rem2}, etc. This type of studies provided with an operational characterization of discord. Consequently such characterization emerged as a useful technique to segregate separable states based on their utility as a resource in these information processing tasks. For instance, in remote state preparation task, any zero discord state was argued to be useless\cite{rem1}. Here, it should be noted out such operational interpretation of discord was argued to be incomplete in \cite{rem2} where authors provided with examples of both useless non zero discord states and useful zero discord states. At this junction, it becomes interesting to explore whether such useless states can again be rendered useful as a resource. Application of global unitary operations\cite{gam} turns out to be useful in this context.\\
Extensive study of global unitary operations over past few years suggests effectiveness of these operations to exploit nonclassicality of some quantum states which otherwise fail to exhibit the same. For instance entanglement can be generated from some separable states\cite{guo1,guo2,guo3} when density matrices of corresponding states are given suitable global basis change via unitary operations\cite{gam}. Extracting Bell nonlocal correlations\cite{guoch} from local states(standard Bell-CHSH scenario) and exploiting asymmetric steering phenomenon from unsteerable quantum states\cite{guost} are also observed when subjected to suitable global unitary operations. Analogous results exist when one considers negative conditional entropy of some family of quantum states\cite{guoen}. In present manuscript we intend to make similar study in ambit of quantum discord.\\
Here we explore the role of global unitary operations to extract positive quantum discord from zero discord states\cite{QD,modi}. For our purpose we have considered general form of a two qubit density matrix after being subjected to nonlocal operation corresponding to an arbitrary global unitary operation\cite{gam}. In literature, the concept of `absoluteness' emerged for characterizing quantum states in context of global unitary operations\cite{guo1,guo2,guo3,guoch,guost,guoen}. Terms such as `absolutely separable'\cite{guo1,guo2,guo3}, `absolutely Bell-CHSH local'\cite{guoch}, `absolutely unsteerable'\cite{guost}, `absolute conditional non-negative von Neumann entropy'\cite{guoen} were framed to categorize quantum states which maintained their characteristic of separability, Bell-CHSH locality, unsteerability and positive conditional entropy respectively even after being subjected to global unitary operations. Following the trend, notion of \textit{absolute zero discord state} emerges in our study. Interestingly, it is observed that given an arbitrary two qubit zero discord state(other than maximally mixed state), it is always possible to obtain a non zero quantum discord state. Consequently the family of of absolute zero discord states turns out to be a singleton set. In course of exploring, we have been able to point out the suitable global unitaries(at least one set) in this context.\\
Rest of our work is organized as follows: In Sec.\ref{mot}, we present our motivation behind this work. In Sec.\ref{pre}, we provide with some pre-requisites. In Sec.\ref{main1}, we present our findings followed by some practical implications in Sec.\ref{prac}. We end our discussion with some concluding remarks in Sec.\ref{conc}.
\section{Motivation}\label{mot}
Broadly speaking, consumption of weak resources such as quantum discord in various `better-than-classical` quantum information processing tasks\cite{eff2,eff4,p2,p3,rem1,rem2} generates the basic idea of this work. Owing to quantum discord's utility as a resource, it will be interesting to explore means via which useless states(in perspective of QIP tasks) can be rendered as a valid resource. Now, as already discussed, there exist evidences of global unitaries to exploit nonclassical trait from quantum states more efficiently than any other standard means\cite{gam}. This generates an obvious intuition of creating positive discord from zero discord states via application of suitable global unitary operations. Such an intuition provides motivation to explore accordingly so as to characterize quantum discord under global basis change. Owing to use of discord as a non-classical resource in QIP tasks, such characterization will therefore be significant for practical purposes.
\section{Mathematical Pre-requisites}\label{pre}
 \subsection{Bloch Vector Representation}
Let $\rho_{\mathcal{AB}}$ denote a two qubit state shared between two parties Alice($\mathcal{A}$) and Bob($\mathcal{B}$) can be represented as:
\begin{equation}\label{st4}
\small{\rho_{\mathcal{AB}}}=\small{\frac{1}{4}(\mathbb{I}_{2\times2}+\vec{\mathfrak{a}}.\vec{\sigma}\otimes \mathbb{I}_2+\mathbb{I}_2\otimes \vec{\mathfrak{b}}.\vec{\sigma}+\sum_{j_1,j_2=1}^{3}\mathfrak{t}_{j_1j_2}\sigma_{j_1}\otimes\sigma_{j_2})},
\end{equation}
with $\vec{\sigma}$$=$$(\sigma_1,\sigma_2,\sigma_3), $ $\sigma_{j_k}$ denoting the Pauli operators lying along three mutually perpendicular directions($j_k$$=$$1,2,3$). $\vec{\mathfrak{a}}$$=$$(l_1,l_2,l_3)$ and $\vec{\mathfrak{b}}$$=$$(r_1,r_2,r_3)$ stand for the local Bloch vectors($\vec{\mathfrak{a}},\vec{\mathfrak{b}}$$\in$$\mathbb{R}^3$) of party $\mathcal{A}$ and $\mathcal{B}$ respectively with $|\vec{\mathfrak{a}}|,|\vec{\mathfrak{b}}|$$\leq$$1$ and $(\mathfrak{t}_{i,j})_{3\times3}$ denotes the correlation tensor $\mathcal{T}$(a real matrix).
 The components $\mathfrak{t}_{j_1j_2}$ are given by $\mathfrak{t}_{j_1j_2}$$=$$\textmd{Tr}[\rho_{\mathcal{AB}}\,\sigma_{j_1}\otimes\sigma_{j_2}].$ \\
 \subsection{Global Unitary Operations}
Global unitary operations, often referred to as general unitary operations\cite{gam}, when applied on $\rho_{\mathcal{AB}},$ corresponds to application of global basis change of the density matrix. Let $\mathcal{U}$ denote a global unitary operation. $\mathcal{U}$ comprises of two parts:
\begin{itemize}
  \item local unitary operations $\mathcal{U}_{\mathcal{A}}$$\otimes$$\mathcal{U}_{\mathcal{B}}:$\\ $\mathcal{U}_{\mathcal{A}},\,\mathcal{U}_{\mathcal{B}}$  correspond to local basis change of subsystems $\mathcal{A}$ and $\mathcal{B}$ respectively. Under application of local unitaries, transformation of $\rho_{\mathcal{AB}}$ to a new density matrix $\rho_{\mathcal{AB}}^{'}$ is characterized by:
\begin{eqnarray}\label{lue1}
\vec{\mathfrak{a}}\rightarrow \vec{\mathfrak{a}^{'}}= \mathcal{Q}_{\mathcal{A}} \vec{\mathfrak{a}}\\
 \vec{\mathfrak{b}}\rightarrow \vec{\mathfrak{b}^{'}}= \mathcal{Q}_{\mathcal{B}} \vec{\mathfrak{b}}\\
 \mathcal{T}\rightarrow \mathcal{T}^{'}= \mathcal{Q}_{\mathcal{A}}\mathcal{T}\mathcal{Q}_{\mathcal{B}}^{\dagger}
\end{eqnarray}
where $\mathcal{Q}_{\mathcal{A}},\,\mathcal{Q}_{\mathcal{B}}$ denote rotation matrices corresponding to unitary operations $\mathcal{U}_{\mathcal{A}},\mathcal{U}_{\mathcal{B}}$ respectively.\\
\item  $\textmd{nonlocal unitary operation}$ $\widehat{\mathcal{U}}$$\equiv$$\widehat{\mathcal{U}}(\phi_1,\phi_2,\phi_3)$$=
    $$\textmd{exp}[\frac{1}{2}\sum_{i=1}^3\phi_i\sigma_i].\,$$\textmd{When subjected}$\\ to nonlocal unitary $\widehat{\mathcal{U}}$, changes in Bloch matrix representation of $\rho_{\mathcal{AB}}$ are given by:
\begin{eqnarray}\label{nlue1}
\mathfrak{a}_k^{'}= \mathfrak{a}_k \cos(\phi_i)\cos(\phi_j)+\mathfrak{b}_k \sin(\phi_i)\sin(\phi_j)+\nonumber\\
\epsilon_{ijk}(\mathfrak{t}_{ij} \cos(\phi_i)\sin(\phi_j)-\mathfrak{t}_{ji}\sin(\phi_i)\cos(\phi_j))\\
\mathfrak{b}_k^{'}= \mathfrak{b}_k \cos(\phi_i)\cos(\phi_j)+\mathfrak{a}_k \sin(\phi_i)\sin(\phi_j)+\nonumber\\
\epsilon_{ijk}(\mathfrak{t}_{ji} \cos(\phi_i)\sin(\phi_j)-\mathfrak{t}_{ij}\sin(\phi_i)\cos(\phi_j))\\
\mathfrak{t}_{ij}^{'}= \mathfrak{t}_{ij} \cos(\phi_i)\cos(\phi_j)+\mathfrak{t}_{ji} \sin(\phi_i)\sin(\phi_j)-\nonumber\\
\epsilon_{ijk}(\mathfrak{a}_{k} \cos(\phi_i)\sin(\phi_j)-\mathfrak{b}_{k}\sin(\phi_i)\cos(\phi_j))
\end{eqnarray}
where $\epsilon_{ijk}$ stand for Levi-Civita symbols\cite{gam}, $i,j,k$$\in$$\{1,2,3\}$ and $i,j,k$ are distinct for first two relations.
\end{itemize}
Application of a general unitary operation $\mathcal{U}$ on $\rho_{\mathcal{AB}}$ thus may be interpreted as applying local unitary operations $\mathcal{U}_{\mathcal{A}}^{1},\mathcal{U}_{\mathcal{B}}^{1}$ on individual subsystems $\mathcal{A}$ and $\mathcal{B}$ respectively followed by nonlocal unitary operation $\widehat{\mathcal{U}}$ on the whole system and then again applying local unitary operations $\mathcal{U}_{\mathcal{A}}^{2},\mathcal{U}_{\mathcal{B}}^{2}$ on $\mathcal{A}$ and $\mathcal{B}$ respectively. If $\rho_{\mathcal{AB}}$$\rightarrow$$\mathcal{U}\rho_{\mathcal{AB}}\mathcal{U}^{\dagger}$$=$$\rho_{\mathcal{AB}}^{'}$(say) denote the transformation under application a general unitary $\mathcal{U}$ then:
 \begin{equation}\label{changed}
    \rho_{\mathcal{\small{AB}}}^{'}
    =\small{\mathcal{U}_{\mathcal{A}}^2\otimes\mathcal{U}_{\mathcal{B}}^2\widehat{\mathcal{U}}}
    \mathcal{U}_{\mathcal{A}}^1\otimes\mathcal{U}_{\mathcal{B}}^1
    \rho_{\mathcal{\small{AB}}}(\mathcal{U}_{\mathcal{A}}^1\otimes
    \mathcal{U}_{\mathcal{B}}^1)^{\dagger}(\widehat{\mathcal{U}})^{\dagger}(\mathcal{U}_{\mathcal{A}}^2\otimes
    \mathcal{U}_{\mathcal{B}}^2)^{\dagger}).
 \end{equation}
 \subsection{Quantum Discord}
Ollivier and Zurek\cite{QD} framed the idea of \textit{Quantum Discord} to measure genuinely quantum correlation content. For any bipartite state $\rho_{\mathcal{AB}},$ total content of correlations is measured by its quantum mutual information:
 \begin{equation}\label{mu1}
I(\mathcal{A}:\mathcal{B})=\mathcal{S}(\mathcal{A})+\mathcal{S}(\mathcal{B})-\mathcal{S}(\mathcal{AB}).
 \end{equation}
 where $\mathcal{S}(\mathcal{X})$$=$$-\textmd{Tr}(\rho_\mathcal{X} \log_2\rho_\mathcal{X})$ is defined to be Von Neumann entropy of a state $\rho_\mathcal{X}$ of system $\mathcal{X}$.
Total classical correlation of $\rho_{\mathcal{AB}}$ is calculated as:
 \begin{equation}\label{cl1}
 \begin{split}
J(\mathcal{B}/\mathcal{A})=\mathcal{S}(\mathcal{B})-\max_{\{\vec{u}^{\mathcal{A}}\}}[\mathcal{S}(\mathcal{B}|\{\Pi_j^{\mathcal{A}}\})]
  \end{split}
 \end{equation}
Here, maximization is made over all possible directions $\vec{u}^{\mathcal{A}}$(unit vector) of complete set ($i$$=$$0,1$ for qubit systems) of orthogonal projections $\{\Pi_{i}^{\mathcal{A}}\}_i$ on Alice's($\mathcal{A}$) subsystem: $ \Pi_{i}^{\mathcal{A}}$$=$$\frac{1}{2}(\mathbb{I}_2+(-1)^i \vec{u}^{\mathcal{A}}.\vec{\sigma})$
with $i$$=$$0,1$ corresponding to two outputs of $\mathcal{A}$'s orthogonal projection along $\vec{u}^{\mathcal{A}}.$ $\mathcal{S}(\mathcal{B}|\{\Pi_i^{\mathcal{A}}\})$$=$$\sum_i p_i\mathcal{S}(\rho_{\small{|i}}^{\mathcal{B}})$ with
$\rho_{\small{|i}}^{\mathcal{B}}(i$$=$$0,1)$ denoting states that Bob($\mathcal{B}$) will obtain conditioned on $\mathcal{A}$ obtaining output $i$ and $p_i$ is the probability with which $\mathcal{A}$ obtains output $i$ while projecting its subsystem along $\vec{u}^{\mathcal{A}},$ $p_i$$=$$\textmd{Tr}[(\Pi_{i}^{\mathcal{A}}\otimes\mathbb{I}_{\small{2}}).\rho_{\mathcal{AB}}](i$$=$$0,1).$
For a bipartite state $\rho_{\mathcal{AB}},\, $quantum discord\cite{QD} is defined as the difference existing between its total correlation content(Eq.(\ref{mu1})) and classical correlation content(Eq.(\ref{cl1})):
 \begin{equation}\label{QD}
  \mathbb{D}(\mathcal{B}/\mathcal{A})=I(\mathcal{A}:\mathcal{B})-J(\mathcal{B}/\mathcal{A}).
 \end{equation}
Quantum discord, in general, is asymmetric in nature\cite{modi}, i.e., $\mathbb{D}(\mathcal{B}/\mathcal{A})\ne \mathbb{D}(\mathcal{A}/\mathcal{B})$.\\
For a bipartite state $\rho_{\mathcal{AB}}$, $\mathbb{D}_{\rho_{\mathcal{AB}}}(\mathcal{B}/\mathcal{A})$(Eq.(\ref{QD})) vanishes if and only if there exists a complete set of orthonormal projectors $\{\Pi_{i}^{\mathcal{A}}\}_i$  having rank $1$ so that $\rho_{\mathcal{AB}}$ takes the form\cite{modi}:
\begin{equation}\label{cq}
    \rho_{\mathcal{AB}}=\sum_{i=0}^1p_i \Pi_{i}^{\mathcal{A}}\otimes\rho_{|i}^{\mathcal{B}}
\end{equation}
These states(Eq.(\ref{cq})) are referred to as \textit{classical-quantum} states in the sense that it is classical with respect to  $\mathcal{A}$ whereas quantum with respect to subsystem $\mathcal{B}.$ Analogously, $\mathbb{D}_{\rho_{\mathcal{AB}}}(\mathcal{A}/\mathcal{B})$ vanishes if and only if $\rho_{\mathcal{AB}}$ is quantum and classical with respect to subsystem $\mathcal{A}$ and $\mathcal{B}$  respectively, i.e., a \textit{quantum-classical} state:
\begin{equation}\label{qc}
    \rho_{\mathcal{AB}}=\sum_{i=0}^1q_i \rho_{|i}^{\mathcal{A}}\otimes\Pi_{i}^{\mathcal{B}}
\end{equation}
where the notations are analogously defined exchanging roles of parties $\mathcal{A}$ and $\mathcal{B}.$\\
For our purpose, we will refer to both $\mathbb{D}_{\rho_{\mathcal{AB}}}(\mathcal{B}/\mathcal{A})$ and  $\mathbb{D}_{\rho_{\mathcal{AB}}}(\mathcal{A}/\mathcal{B})$ as `one-way` discord\cite{self}. If $\mathbb{D}_{\rho_{\mathcal{AB}}}(\mathcal{B}/\mathcal{A})$ and $\mathbb{D}_{\rho_{\mathcal{AB}}}(\mathcal{A}/\mathcal{B})$ both turn out to be positive for $\rho_{\mathcal{AB}},$ then $\rho_{\mathcal{AB}}$ will be referred to as a `both-way' discord state\cite{self}.\\
\subsection{Classical-Quantum States and Quantum-Classical States Under Local Unitary Operations}
Let local unitary operations be applied on a classical-quantum state $\rho_{\mathcal{AB}}.$ All possible forms of the transformed state were derived in \cite{self}. Those forms are enlisted in Table.\ref{table:ta1}. Similarly transformed form(under local unitary operations) of a quantum-classical state\cite{self} are given in Table.\ref{table:ta2}.\\
\begin{center}
\begin{table}[htp]
\caption{Under local unitary operations, any classical-quantum state(Eq.(\ref{cq})) will assume one of these  possible forms\cite{self}. So if $\rho_{\mathfrak{AB}}^{(1)}$(Eq.(\ref{ist})), obtained after local unitary operations, is classical-quantum then its state parameters will be characterized by one of the forms enlisted here. $\vec{w}^a$ indicates the direction of orthogonal projectors $\Pi^{\mathcal{A}}_i$(Eq.(\ref{cq})). $\mathfrak{s}_{11},\mathfrak{s}_{22},\mathfrak{s}_{33}$ denote diagonal elements of the correlations tensor $\mathcal{S}$ obtained from correlation tensor $\mathcal{T}$ of the given state $\rho_{\mathcal{AB}}$ after application of local unitary operations   }
\begin{center}
\begin{tabular}{|c|c|}
\hline
State&$\vec{w}^a$\\
\hline
  $\frac{1}{4}(\mathbb{I}_{2\times2}+\vec{\mathfrak{m}}.\vec{\sigma}\otimes \mathbb{I}_2+$& $\pm\frac{1}{\sqrt{\sum _{j=1}^3 a_j^2}}(a_1,a_2,a_3)$\\
  $\mathbb{I}_2\otimes \vec{  \mathfrak{n}}.\vec{\sigma})$&\,\\
\hline
$\frac{1}{4}(\mathbb{I}_{2\times2}+\mathfrak{m}_i\sigma_i\otimes \mathbb{I}_2+$& $\vec{w}^a$ has $w^a_j$$=$$\pm 1(j$$=$$1,2,3)$ \\
$\mathbb{I}_2\otimes \vec{\mathfrak{n}}.\vec{\sigma}+\mathfrak{s}_{ii}\sigma_{i}\otimes\sigma_{i})(i$$=$$1,2,3)$&\small{with other two components zero.}\\
\hline
\end{tabular}\\
\end{center}
\label{table:ta1}
\end{table}
\end{center}
\begin{center}
\begin{table}[htp]
\caption{If $\rho_{\mathfrak{AB}}^{(1)}$(Eq.(\ref{ist})) is a quantum-classical state(Eq.(\ref{qc})), then it will take one of the possible forms given below. $\vec{w}^b$ indicates the direction of orthogonal projectors $\Pi^{\mathcal{B}}_i$(Eq.(\ref{qc})).  }
\begin{center}
\begin{tabular}{|c|c|}
\hline
State&$\vec{w}^b$\\
\hline
  $\frac{1}{4}(\mathbb{I}_{2\times2}+\vec{\mathfrak{m}}.\vec{\sigma}\otimes \mathbb{I}_2+$& $\pm\frac{1}{\sqrt{\sum _{j=1}^3 b_j^2}}(b_1,b_2,b_3)$\\
  $\mathbb{I}_2\otimes \vec{
  \mathfrak{n}}.\vec{\sigma})$&\,\\
\hline
$\frac{1}{4}(\mathbb{I}_{2\times2}+\vec{\mathfrak{m}}.\vec{\sigma}\otimes\mathbb{I}_2 +$& $\vec{w}^b$ has $w^b_j$$=$$\pm 1(j$$=$$1,2,3)$  and \\
$\mathbb{I}_2\otimes \mathfrak{n}_i\sigma_i+\mathfrak{s}_{ii}\sigma_{i}\otimes\sigma_{i})(i$$=$$1,2,3)$&other two components zero.\\
\hline
\end{tabular}\\
\end{center}
\label{table:ta2}
\end{table}
\end{center}
\subsection{Remote State Preparation(RSP)}
It is a bipartite quantum information processing task\cite{rem3,rem4} in which one of the parties, say Alice remotely prepares a state in the location of other party, say Bob. So in this task, Alice aims in communicating a known quantum state to Bob without performing any quantum communication. This task is therefore almost similar to that of quantum teleportation\cite{tele1}. In teleportation, however, the sender has no knowledge about the state to be sent to the receiver\cite{tele1}. Below we briefly review the task of remote state preparation.\\
\textit{Protocol:} Let a singlet $  |\psi^{-}\rangle$$=$$\frac{|01\rangle-|10\rangle}{\sqrt{2}}$ be shared between the two parties.
Let Alice aims to prepare an equatorial state vector $|\phi\rangle$$=$$\frac{|0\rangle+\exp(\imath \theta)|1\rangle}{\sqrt{2}}$ in Bob's lab. She performs orthogonal projective measurement on her part of the singlet in the basis $\{|\phi\rangle,|\phi^{\bot}\rangle\}.$ She communicates a cbit $0$ or $1$(say) on receiving $|\phi\rangle$ or $|\phi^{\bot}\rangle$ respectively. As they share a singlet, on receiving $1,$ Bob knows that Alice has obtained $|\phi^{\bot}\rangle$ and consequently his subsystem(part of the singlet in his lab) corresponds to $|\phi\rangle.$ If he receives $0,$ he knows that his subsystem now corresponds to $|\phi^{\bot}\rangle.$ He then performs a $\sigma_3$(rotating his subsystem by $180^{\circ}$ so as to get the required qubit $|\phi\rangle$. So at the end of the protocol, Alice succeeds in preparing remotely the qubit $|\phi\rangle$ in Bob's lab. For the protocol, bipartite entanglement(singlet) and one cbit communication are used as resources unlike that in teleportation where two cbits of communication are required.
\section{Characterizing Quantum discord under global unitary operations}\label{main1}
Here we analyze the role of global unitary operations to exploit non vanishing quantum discord from zero discord states. This in turn characterize states which can be converted into resources under suitable global basis change. Now, quantum discord being asymmetric in nature, we treat `one-way' discord $\mathbb{D}_{\rho_{\mathcal{AB}}}(\mathcal{B}/\mathcal{A})$ and $\mathbb{D}_{\rho_{\mathcal{AB}}}(\mathcal{A}/\mathcal{B})$ simultaneously. Let the set of all classical-quantum(Eq.(\ref{cq})) and quantum-classical(Eq.(\ref{qc})) states be denoted as:
\begin{equation}\label{cqs}
    CQ=\{\sigma_{\mathcal{AB}}:\mathbb{D}_{\sigma_{\mathcal{AB}}}(\mathcal{B}/\mathcal{A})=0\}
\end{equation}
\begin{equation}\label{cqs}
    QC=\{\sigma_{\mathcal{AB}}:\mathbb{D}_{\sigma_{\mathcal{AB}}}(\mathcal{A}/\mathcal{B})=0\}
\end{equation}
\subsection{Absolute Classical-Quantum States}
$\textmd{For\,characterizing}\,`\textmd{one-way}'\,\textmd{quantum\,discord}$\\
$\mathbb{D}_{\sigma_{\mathcal{AB}}}(\mathcal{B}/\mathcal{A})$ under global unitary operation we first define absolute classical-quantum states:\\
\textit{\textbf{Definition:}}\textit{Any classical-quantum state whose discord remains invariant after application of arbitrary global unitary operation $\mathcal{U}$ is said to be an absolute classical-quantum state.}\\
If $ACQ$ denote the set of absolute classical-quantum states then,
\begin{equation}\label{acq}
    ACQ=\{\sigma_{\mathcal{AB}}\in CQ:\mathbb{D}_{\mathcal{U}\sigma_{\mathcal{AB}}\mathcal{U}^{\dagger}}(\mathcal{B}/\mathcal{A})=0\},
\end{equation}
where effect of global unitary operation $\mathcal{U}$ is given by Eq.(\ref{changed}). Given an arbitrary classical-quantum state($\rho_{\mathcal{AB}}$), if there exist at least one suitable global unitary operation for which the transformed state $\rho_{\mathcal{AB}}^{'}$ has non vanishing `one-way' discord, then $\rho_{\mathcal{AB}}$ is not a member of $ACQ.$ Clearly, for effectiveness of applying global unitary operations to exploit `one-way' discord of an arbitrary two qubit state $\rho_{\mathcal{AB}}$, one requires
$\mathbb{D}_{\mathcal{U}\sigma_{\mathcal{AB}}\mathcal{U}^{\dagger}}(\mathcal{B}/\mathcal{A})$$>$$0.$ This in turns requires:
\begin{equation}\label{entropic2}
    \mathcal{S}((\mathcal{A})^{'})+\max_{\{(\vec{u}^{\mathcal{A}})\}}[\mathcal{S}((\mathcal{B}|\{\Pi_j^{\mathcal{A}}\})^{'})]
    -\mathcal{S}((\mathcal{AB})^{'})>0
\end{equation}
where $(.)^{'}$ is used to denote components related to the transformed state $\rho_{\mathcal{AB}}^{'}$(Eq.(\ref{changed})). Owing to complexity due to involvement of maximization in the entropic constraint(Eq.(\ref{entropic2})), we adopt an alternative approach to characterize ACQ. Bloch vector representation of $\rho_{\mathcal{AB}}$ after application of nonlocal unitary operations $\widehat{\mathcal{U}}$ serves our purpose. \\
\textbf{Theorem 1:} \textit{Set of absolute classical-quantum states is singleton. In particular, $ACQ$$=$$\{\frac{1}{4}\mathbb{I}_{2\times 2}\}.$ }\\
\textit{Proof:} Let $\rho_{\mathcal{AB}}$$\in$$CQ.$ Let it be subjected to an arbitrary general(global) unitary operation $\mathcal{U}.$ It has already been discussed before that application of a general(global) unitary $\mathcal{U}$ over any two qubit state can be interpreted as that of applying local unitary operations(on subsystems) followed by nonlocal unitary operations on the whole system and then again followed by local unitary operations. Let $\rho_{\mathcal{AB}}^{(1)},\,\rho_{\mathcal{AB}}^{(2)},\,\rho_{\mathcal{AB}}^{(3)}$ denote the transformed states in subsequent stages of transformation $\rho_{\mathcal{AB}}$$\rightarrow$$\rho_{\mathcal{AB}}^{'}$:
\begin{eqnarray}\label{ist}
    \rho_{\mathcal{AB}}^{(1)}=\mathcal{U}_{\mathcal{A}}^1\otimes \mathcal{U}_{\mathcal{B}}^1 \rho_{\mathcal{AB}}(\mathcal{U}_{\mathcal{A}}^1\otimes \mathcal{U}_{\mathcal{B}}^1)^{\dagger}\nonumber\\
  \rho_{\mathcal{AB}}^{(2)}=\widehat{\mathcal{U}} \rho_{\mathcal{AB}}^{(1)}(\widehat{\mathcal{U}})^{\dagger}\:\qquad\qquad\qquad\nonumber\\
     \rho_{\mathcal{AB}}^{(3)}=\rho_{\mathcal{AB}}^{'}=\mathcal{U}_{\mathcal{A}}^2\otimes \mathcal{U}_{\mathcal{B}}^2 \rho_{\mathcal{AB}}^{(2)}(\mathcal{U}_{\mathcal{A}}^2\otimes \mathcal{U}_{\mathcal{B}}^2)^{\dagger}
\end{eqnarray}
Now applying local unitaries has no effect on quantum discord of a two qubit state\cite{modi}. So if $\rho_{\mathcal{AB}}$$\in$$CQ,$ then $\rho_{\mathcal{AB}}^{(1)}$$\in$$CQ.$ Now as $\rho_{\mathcal{AB}}^{(1)}$ is a classical-quantum state(Eq.(\ref{cq})) so all possible forms(Bloch vector representation\cite{self}) of $\rho_{\mathcal{AB}}^{(1)}$ are given by Table.\ref{table:ta1}.
Nonlocal unitary operation $\widehat{\mathcal{U}}$ is now applied on $\rho_{\mathcal{AB}}^{(1)}.$ The detailed analysis of applying $\widehat{\mathcal{U}}$ on all possible forms of $\rho_{\mathcal{AB}}^{(1)}$ is given in Appendix.A
which shows that for every possible form of $\rho_{\mathcal{AB}}^{(1)}$ except $\frac{1}{4}\mathbb{I}_{2\times 2},$ there exists a nonlocal unitary operation(see Table.\ref{table:ta3} for suitable values of parameters $\phi_1,\phi_2,\phi_3$) such that resulting state $\rho_{\mathcal{AB}}^{(2)}$
is not a classical-quantum state. Now $\mathbb{D}(\mathcal{B}/\mathcal{A})$$\neq $$0$ if and only if $\rho_{\mathcal{AB}}^{(2)}$ is not a classical-quantum state\cite{modi}. Hence every $\rho_{\mathcal{AB}}^{(2)}$ except $\frac{1}{4}\mathbb{I}_{2\times 2}$ has non zero `one-way' discord($\mathbb{D}(\mathcal{B}/\mathcal{A})$$\neq$$0.$
Lastly local unitary operation $\mathcal{U}_{\mathcal{A}}^2\otimes \mathcal{U}_{\mathcal{B}}^2$ is applied resulting in state $\rho_{\mathcal{AB}}^{(3)}.$ $\mathbb{D}(\mathcal{B}/\mathcal{A})$ remaining invariant under local unitaries, any possible form of
$\rho_{\mathcal{AB}}^{(3)}$ except $\frac{1}{4}\mathbb{I}_{2\times 2}$ has non zero `one-way' discord$.\,\textmd{Hence},$ excepting the maximally mixed state($\frac{1}{4}\mathbb{I}_{2\times 2}$), any member $\rho_{\mathcal{AB}}$ from the set of classical-quantum states($CQ$) gets transformed into a`one-way' non zero discord state $\rho_{\mathcal{AB}}^{'}.$ Consequently $ACQ$$=$$\{\frac{\mathbb{I}_{2\times 2}}{4}\}$$\blacksquare$\\
Discord, being asymmetric, we now give a formal characterization of absolute quantum-classical states separately.\\
\subsection{Absolute Quantum-Classical States}
\textit{\textbf{Definition:}} \textit{Any quantum-classical state whose discord remains invariant after application of arbitrary global unitary operation $\mathcal{U}$ is said to be an absolute quantum-classical state.}\\
Let $AQC$ denote the set of absolute quantum-classical states:
\begin{equation}\label{aqc}
    AQC=\{\sigma_{\mathcal{AB}}\in QC:\mathbb{D}_{\mathcal{U}\sigma_{\mathcal{AB}}\mathcal{U}^{\dagger}}(\mathcal{A}/\mathcal{B})=0\},
\end{equation}
Effect of global unitary operation $\mathcal{U}$ is given by Eq.(\ref{changed}). Analogous to classical-quantum states, we next characterize $AQC$.\\
\textbf{Theorem 2:} \textit{Set of absolute quantum-classical states is singleton having maximally mixed state as its only member: $AQC$$=$$\{\frac{1}{4}\mathbb{I}_{2\times 2}\}.$ }\\
\textit{Proof:} Proof is similar to that of theorem.1, with now applying global unitary operation on an arbitrary quantum-classical state  $\rho_{\mathcal{AB}}.$ Action of nonlocal unitary operation $\widehat{\mathcal{U}}$(Table.\ref{table:ta4}) on every possible form of quantum-classical state, obtained after applying $\mathcal{U}_{\mathcal{A}}^{(1)}\otimes \mathcal{U}_{\mathcal{B}}^{(1)}$(Table.\ref{table:ta2}) is discussed in Appendix.B.\\
Characterization of $ACQ$ and $AQC,$ facilitates to analyze role of global unitaries over `both-way' zero discord states\cite{self}.
\subsection{Absolute Zero Discord States}
As mentioned in Sec.\ref{pre},  any two qubit state $\rho_{\mathcal{AB}}$ for which $\mathbb{D}_{\rho_{\mathcal{AB}}}(\mathcal{B}/\mathcal{A})$$=$$\mathbb{D}_{\rho_{\mathcal{AB}}}(\mathcal{A}/\mathcal{B})$$=$$0,$ is said to be a `both-way' zero discord, or simply \textit{zero discord} state. Here we characterize such states in present context.\\
\textit{\textbf{Definition:}} \textit{Any two qubit state which remains zero discord state after application of arbitrary global unitary operation $\mathcal{U}$ is said to be an absolute zero discord state.}\\
Let $AZD$ denote collection of all absolute zero discord states:
\begin{equation}\label{azd}
   \small{ AZD}=\{\small{\sigma_{\mathcal{AB}}\in ZD:\mathbb{D}_{\mathcal{U}\sigma_{\mathcal{AB}}\mathcal{U}^{\dagger}}(\mathcal{A}/\mathcal{B})}=
    \mathbb{D}_{\mathcal{U}\sigma_{\mathcal{AB}}\mathcal{U}^{\dagger}}(\mathcal{B}/\mathcal{A})=0\},
\end{equation}
where $ZD$ denote the set of all zero discord states. Arguments made so far implies that $AZD$ also contains one element.\\
\textbf{Theorem 3:} \textit{Set of absolute zero discord states($AZD$) contains only maximally mixed state.}\\
\subsection{Illustrations}\label{illus}
Observing effectiveness of global unitary operations to convert zero discord state to non zero discord state, we proceed to discuss the mechanism of this conversion in details.\\
\textit{Application of Global Unitary Operations:} Given an arbitrary two qubit state $\rho_{\mathcal{AB}},$ with correlation tensor $\mathcal{T},$ singular value decomposition of $\mathcal{T}$ may be obtained by performing suitable local unitary operations $\mathcal{U}_{\mathcal{A}}^1\otimes \mathcal{U}_{\mathcal{B}}^1$ over $\rho_{\mathcal{AB}}$\cite{gam}. Let $\rho_{\mathcal{AB}}$ be a classical-quantum state. Let $\kappa_i^{L}(i$$=$$1,2,3)$ and $,\kappa_i^{R}(i$$=$$1,2,3)$ denote orhonormalized left and right singular vectors of $\mathcal{T}$ respectively. $\forall i,$ denoting $\kappa_i^{L(R)}$$\in$$\mathbb{R}^3$ as $(\kappa_{i1}^{L(R)},\kappa_{i2}^{L(R)},\kappa_{i3}^{L(R)}),$ local unitary matrices $\mathcal{U}_{\mathcal{A}}^1,\mathcal{U}_{\mathcal{B}}^1$ are given by:
\begin{eqnarray}\label{uni1}
\mathcal{U}_{\mathcal{A}}^1=\left(\begin{array}{ccc}
      \kappa_{11}^{\small{L}} &\kappa_{12}^{\small{L}} &\kappa_{13}^{\small{L}} \\
      \,&\,&\,\\
       \kappa_{21}^{\small{L}} &\kappa_{22}^{\small{L}} &\kappa_{23}^{\small{L}} \\
       \,&\,&\,\\
         \kappa_{31}^{\small{L}} &\kappa_{32}^{\small{L}} &\kappa_{33}^{\small{L}} \\
         \,&\,&\,\\
       \end{array} \right)\nonumber\\
\mathcal{U}_{\mathcal{B}}^1=\left(\begin{array}{ccc}
      \kappa_{11}^{\small{R}} &\kappa_{12}^{\small{R}} &\kappa_{13}^{\small{R}} \\
      \,&\,&\,\\
       \kappa_{21}^{\small{R}} &\kappa_{22}^{\small{R}} &\kappa_{23}^{\small{R}} \\
       \,&\,&\,\\
         \kappa_{31}^{\small{R}} &\kappa_{32}^{\small{R}} &\kappa_{33}^{\small{R}} \\
       \end{array} \right)
\end{eqnarray}
$\rho_{\mathcal{AB}},$ being a classical-quantum state, after application of the local unitary operations $\mathcal{U}_{\mathcal{A}}^1\otimes \mathcal{U}_{\mathcal{B}}^1$(Eq.(\ref{uni1})), $\rho_{\mathcal{AB}}^{(1)}$(Eq.(\ref{ist})) corresponds to one of the possible forms prescribed in Table.\ref{table:ta1}. Then observing the exact form of $\rho_{\mathcal{AB}}^{(1)}$ the suitable nonlocal unitary operation $\widehat{\mathcal{U}}(\phi_1,\phi_2,\phi_3)$ to be applied is chosen from Table.\ref{table:ta3}(which enlists required nonlocal unitary for every possible form of classical-quantum state given in Table.\ref{table:ta1}). \\
Now to obtain `one-way' zero discord state starting from an arbitrary quantum-classical state $\rho_{\mathcal{AB}}$, analogous treatment is to be made with now considering Table.\ref{table:ta2} and Table.\ref{table:ta4} instead of Table.\ref{table:ta1} and Table.\ref{table:ta3} for obvious reasons.\\
\textit{An Example:} Consider a two qubit product state:
\begin{equation}\label{prod}
    \rho_{prod}=\frac{1}{4}(\mathbb{I}_2+\vec{r}^a.\vec{\sigma})\otimes(\mathbb{I}_2+\vec{r}^b.\vec{\sigma})
\end{equation}
where $\vec{r}^{a(b)}$$=$$(r_1^{a(b)},r_2^{a(b)},r_3^{a(b)}).$
$$\,$$
$\mathbb{D}_{\rho_{prod}}(\mathcal{B}/\mathcal{A})$$=$$\mathbb{D}_{\rho_{prod}}(\mathcal{A}/\mathcal{B})$$=$$0$\cite{self}.  So $\rho_{prod}$ is a zero discord state. Correlation tensor is given by:
\begin{equation}\label{ten}
    \mathcal{T}_{prod}= \left(\begin{array}{ccc}
     r_1^{a}r_1^{b} & r_1^{a}r_2^{b} & r_1^{a}r_3^{b} \\
        r_2^{a}r_1^{b} & r_2^{a}r_2^{b} & r_2^{a}r_3^{b}\\
          r_3^{a}r_1^{b} & r_3^{a}r_2^{b} & r_3^{a}r_3^{b} \\
       \end{array} \right)
\end{equation}
The suitable local unitary operations are:
\begin{equation}\label{uni2}
\mathcal{U}_{\mathcal{A}(\mathcal{B})}^1=\left(\begin{array}{ccc}
      -\frac{r_3^{a(b)}}{r_1^{a(b)}n_1^{a(b)}}&0&\frac{1}{n_1^{a(b)}} \\
      \,&\,&\,\\
         -\frac{r_2^{a(b)}}{r_1^{a(b)}n_2^{a(b)}}&\frac{1}{n_2^{a(b)}}&0 \\
         \,&\,&\,\\
          \frac{r_1^{a(b)}}{r_3^{a(b)}n_3^{a(b)}}&\frac{r_2^{a(b)}}{r_3^{a(b)}n_3^{a(b)}}&\frac{1}{n_3^{a(b)}} \\
       \end{array} \right)\nonumber\\
\end{equation}
where $n_1^{a(b)}$$=$$\sqrt{1+(\frac{r_3^{a(b)}}{r_1^{a(b)}})^2},$ $n_2^{a(b)}$$=$$\sqrt{1+(\frac{r_2^{a(b)}}{r_1^{a(b)}})^2}$ and $n_3^{a(b)}$$=$$\sqrt{\frac{(r_1^{a(b)})^2+(r_2^{a(b)})^2+(r_3^{a(b)})^2}{(r_3^{a(b)})^2}}.$ After application of these local unitary operations $\rho_{prod}^{(1)}$ is given by:
\begin{equation}\label{uni3}
\small{\rho_{prod}^{(1)}}=\small{\frac{1}{4}(\mathbb{I}_{2\times2}+\vec{\mathfrak{a}}^{(1)}.\vec{\sigma}\otimes \mathbb{I}_2+\mathbb{I}_2\otimes \vec{\mathfrak{b}}^{(1)}.\vec{\sigma}+\sum_{j_1,j_2=1}^{3}\mathfrak{t}_{j_1j_2}^{(1)}\sigma_{j_1}\otimes\sigma_{j_2})},
\end{equation}
where $\vec{\mathfrak{a}}^{(1)}$$=$$(0,0,n_3^{a}r_3^{a}),$ $\vec{\mathfrak{b}}^{(1)}$$=$$(0,0,n_3^{b}r_3^{b})$ and correlation tensor $\mathcal{T}_{prod}^{(1)}$ is a diagonal matrix $diag(0,0,r_3^ar_3^bn_3^an_3^b).$ Clearly $\rho_{prod}^{(1)}$(Eq.(\ref{uni3})) corresponds to a form in Table.\ref{table:ta1} and also in Table.\ref{table:ta2}. Observing the particular form in Table.\ref{table:ta1}, on application of nonlocal unitary operation $\widehat{\mathcal{U}}(\pi,\frac{\pi}{2},\pi)$(as prescribed by Table.\ref{table:ta3}), resulting state $\rho_{prod}^{(2)}$ is no longer a classical-quantum state($\mathbb{D}_{\rho_{prod}^{(2)}}(\mathcal{B}/\mathcal{A})$$>$$0$). Again treating $\rho_{prod}^{(1)}$ as a quantum-classical state(Table.\ref{table:ta2}), $\widehat{\mathcal{U}}(\pi,\frac{\pi}{2},0)$ is the suitable nonlocal unitary operation(Table.\ref{table:ta4}) application of which gives $\mathbb{D}_{\rho_{prod}^{(2)}}(\mathcal{A}/\mathcal{B})$$>$$0.$ So $\rho_{prod}^{(2)}$ turns out to be a non zero discord state. Lastly, suitable local unitaries $\mathcal{U}_{\mathcal{A}}^2,$ $\mathcal{U}_{\mathcal{B}}^2$ may again be applied so as to obtain a simplified version of the state.  \\
Alternatively, given any $\rho_{\mathcal{AB}}$ one may directly apply suitable nonlocal unitary operation $\widehat{\mathcal{U}}(\phi_1,\phi_2,\phi_3)$ considering $\mathcal{U}_{\mathcal{A}}^1$$=$$\mathcal{U}_{\mathcal{B}}^1$$=$$\frac{\mathbb{I}}{2}.$ One such instance is cited in next section.\\
\textmd{Till date}, absoluteness analyzed with respect to separability\cite{guo1}, Bell-CHSH locality\cite{guoch}, unsteerability\cite{guost}, positive conditional entropy\cite{guoen} has generated families of bipartite quantum states over which global unitary operations are ineffective(in the sense of rendering nonclassical correlations). This in turn marks the failure of global unitary operations to exploit nonclassicality in corresponding perspectives. However, in contrast to existing results, our observations guarantee success of global unitaries over the entire collection of two qubit states leaving aside trivially the maximally mixed state.\\
\section{Practical Implications}\label{prac}
Based on extensive research activities in field of quantum information processing quantum discord emerges as a potential resource\cite{modi}. Some of these tasks are remote state preparation\cite{rem1,rem2,rem3,rem4}, state merging\cite{p2}, quantum dense coding\cite{p2}, restricted quantum gates\cite{restricted gates}, DQC1 models\cite{p8,p10}, noisy states involved quantum metrology protocols\cite{metrology noisy}, quantum  phase transitions\cite{qpt1,qpt2} and many more. So, from practical viewpoint, given any two qubit state $\varrho$, it will be interesting to analyze whether it has non zero quantum discord or not such that it can be used as a resource in a `better than classical' quantum information task. In case, $\varrho$ turns out to be a zero discord state, obviously one may explore whether there exist any other possible way so as to render this useless state as useful resource. Our discussion so far reveals utility of applying global unitary operations in this perspective. To be precise, being capable of transforming a classical-quantum(or quantum-classical) state to at least `one-way' discord state, suitable global unitary operations may be implemented in practical scenarios. Here we consider the task of remote state preparation for further discussion.\\
\subsection{Global Unitaries in Remote State Preparation}
As discussed in sec.\ref{pre}, to perform this task, the two parties share maximal entanglement in ideal situations\cite{rem3}. However, in non ideal situations, any mixed state having non zero `one-way' quantum discord $\mathbb{D}(\mathcal{A}/\mathcal{B})$(in case Alice and Bob are sender and receiver respectively) or classical correlations also suffices for the purpose\cite{rem5}. But then Bob receives the required quantum state with a reduced fidelity $\mathcal{F}$\cite{rem5}. In \cite{rem1}, the authors showed that non zero discord state($\mathbb{D}(\mathcal{A}/\mathcal{B})$$>$$0$) is necessary for RSP protocol. They gave an operational characterization of non zero discord state via this protocol. For their purpose they however used the notion of geometric discord($\mathbb{D}^{(G)}(\mathcal{A}/\mathcal{B})$\cite{alter}):
\begin{equation}\label{opec1}
    \mathcal{F}=(\mathbb{D}^{(G)}(\mathcal{A}/\mathcal{B}))^2=\frac{1}{2}(\lambda_3+\lambda_2)
\end{equation}
where $\lambda_1$$\geq$$\lambda_2$$\geq$$\lambda_3$ denote the eigen values of $\mathcal{T}^{T}\mathcal{T}.$\\
Now, let Alice and Bob have access to only a two qubit zero discord quantum state $\varrho_{\mathcal{AB}}$(say) other than maximally mixed state. Such a state($\varrho_{\mathcal{AB}}$) cannot be used for RSP as fidelity reduces to zero(Eq.(\ref{opec1})). To accomplish the task of remote state preparation with $\varrho_{\mathcal{AB}}$ only, the protocol needs to be modified as follows.\\
\begin{figure}[htb]
\centering
\includegraphics[width=3in]{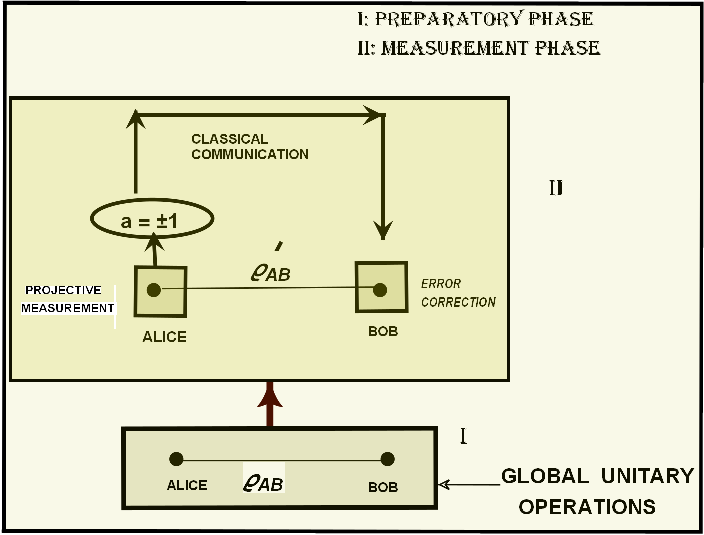}\\
\caption{\emph{Schematic diagram of the modified protocol. $\varrho_{\mathcal{AB}}$ and $\varrho_{\mathcal{AB}}^{'}$ denote the resource state in the preparatory phase and measurement phase respectively. Global unitary operations are performed in preparatory phase whereas standard remote state preparation mechanism\cite{rem3,rem5} constitutes the measurement phase of the protocol. So in this phase Alice first makes projective measurement on her subsystem(of $\varrho_{\mathcal{AB}}^{'}$) and communicates her result to Bob who then corrects his subsystem depending on the communicated bit. }}
\label{fig2}
\end{figure}
\textit{Modified Protocol}: It now consists of two phases: \textit{preparatory phase} followed by \textit{measurement phase}see Fig.\ref{fig2}). \\
In the preparatory phase Alice and bob receive their respective particles constituting $\varrho_{\mathcal{AB}}.$ The state($\varrho_{\mathcal{AB}}$) being known to them, they perform suitable global unitary operations on it so that a new state $\varrho_{\mathcal{AB}}^{'}$ is generated for which $\mathbb{D}_{\varrho_{\mathcal{AB}}^{'}}(\mathcal{A}/\mathcal{B})$ and consequently$\mathbb{D}_{\varrho_{\mathcal{AB}}^{'}}^{(G)}(\mathcal{A}/\mathcal{B})$$>$$0$\cite{rem2}. This ends the preparatory phase. $\varrho_{\mathcal{AB}}^{'}$ can now be used for the purpose of remote state preparation\cite{rem3,rem5} which constitutes the measurement phase. Here Alice and Bob perform local measurements(along with one-way classical communication from Alice to Bob) as done in a standard RSP protocol. Thus Alice succeeds in preparing a quantum state with non zero fidelity remotely in Bobs's lab though they had access to a zero discord state($\varrho_{\mathcal{AB}}$). One may note that the successful accomplishment of the modified protocol relies completely on existence of suitable global unitary operations. That Alice and Bob will definitely find some global unitary for which $\mathbb{D}_{\varrho_{\mathcal{AB}}^{'}}(\mathcal{A}/\mathcal{B})$$>$$0$ is ensured by theorem.2. Also, in preparatory phase, once $\varrho_{\mathcal{AB}}$ can be simplified to one of quantum-classical state forms as given in Table.\ref{table:ta2}(via local basis change), then Alice and Bob know at least one effective nonlocal unitary operation(via Table.\ref{table:ta4}) that may be given. \\
Now, in this context it should be pointed out that in \cite{rem2}, the authors criticized the operational interpretation of discord made in \cite{rem1} saying it to be incomplete. They argued that such a relation(Eq,(\ref{opec1})) may be due to some non optimized version of RSP protocol. As already discussed before, in support of their argument, the authors cited an example of useful zero discord states and also that of a useless non zero discord state. The latter one is relevant with present discussion. In particular, the state(density matrix formalism) cited in \cite{rem2} is as follows:
\begin{equation}\label{opec2}
   \Omega=
   \left(\begin{array}{cccc}
      0.2 &0.1 & 0.1 & 0\\
      0.1 &0.1 & 0 & 0.1\\
      0.1 &0 & 0.3 & 0.1\\
      0 &0.1 & 0.1 & 0.4\\
       \end{array} \right)
\end{equation}
It was reported that $\mathbb{D}_{\Omega}(\mathcal{A}/\mathcal{B})$$=$$0.26$ whereas $\mathbb{D}_{\Omega}^{(G)}(\mathcal{A}/\mathcal{B})$$=$$0.01.$ Consequently fidelity $\mathcal{F}$(Eq.(\ref{opec1})) vanishes($\approx 10^{-4}$). Now $\Omega$(Eq.(\ref{opec2})) can be used for remote state preparation with non zero fidelity via our modified protocol. \\
For that in the preparatory phase let Alice nd Bob perform global unitary operation $\widehat{\mathcal{U}}(\frac{\pi}{2},0,\frac{\pi}{2}).$ $\Omega^{'}$ resulting at the end of the preparatory phase has geometric discord(Eq.(\ref{opec1})) $\mathbb{D}^{(G)}_{\Omega^{'}}(\mathcal{A}/\mathcal{B})$$=$$0.282843.$ Consequently the fidelity $\mathcal{F}$ of the remote state prepared in Bob's lab at end of measurement phase is $0.08.$\\
Modified remote state preparation protocol in turn gives an operational interpretation of global unitary operations.\\
\textit{Operational Interpretation:} Owing to the effectiveness of nonlocal unitary operations to generate `one-way' non zero discord state starting from a quantum-classical state, global unitary operations can be analyzed in perspective of remote state preparation task. Let us assume that a known quantum-classical state $\Phi$ be the only quantum resource shared between Alice and Bob. Alice has  to remotely prepare a quantum state with $\mathcal{F}$$\neq$$0,$(of order $10^{-2}$,say) in Bob's lab. In this context, global unitary operations may be considered as some suitable operations required to be imposed over the supplied bipartite state(as discussed in(\ref{illus})) in the preparatory phase of the modified protocol(Fig.\ref{fig2}) so that the resultant state can then be used in the measurement phase successfully.\\
For a specific instance, consider the quantum-classical state(Table.\ref{table:ta2}):
\begin{equation}\label{opec3}
   \Phi_{\mathcal{AB}}=\small{\frac{1}{4}(\mathbb{I}_{2\times2}+\mathbb{I}_2\otimes \vec{\mathfrak{n}}.\vec{\sigma})},
\end{equation}
where $\vec{\mathfrak{n}}$$=$$(n_1,n_2,n_3)$$\in$$\mathbb{R}^3$  With $||\vec{\mathfrak{n}}||$ denoting length of $\vec{\mathfrak{n}},$ let us discuss the possible cases(depending on values of $n_1,n_2,n_3$) for each of which the modified protocol runs successfully with $\mathcal{F}$$=$$0.01125.$  \\
\begin{itemize}
  \item $||\vec{\mathfrak{n}}||$$\geq$$0.3$: Here $\widehat{\mathcal{U}}(\frac{\pi}{4},\frac{3\pi}{4},\frac{3\pi}{4})$ serves the purpose.
  \item $||\vec{\mathfrak{n}}||$$<$$0.3$ and $n_3$$\geq$$0.3$: $\widehat{\mathcal{U}}(\frac{3\pi}{4},\frac{3\pi}{4},\frac{\pi}{2})$ is useful.
  \item $||\vec{\mathfrak{n}}||$$<$$0.3,$ $n_3$$<$$0.3$ and $n_1$$\geq$$0.3$: $\widehat{\mathcal{U}}(\frac{\pi}{4},\frac{\pi}{4},\frac{\pi}{2})$ is useful.
\end{itemize}
However, for some $\vec{\mathfrak{n}}$ the protocol does not run successfully with $\mathcal{F}$$\geq$$10^{-2}.$  If state $\Phi_{\mathcal{AB}}$(Eq.(\ref{opec3})) characterized by $n_1$<$0.3,$ $n_2$$=$$n_3$$=$$0,$ is supplied, for any $\widehat{\mathcal{U}}(\phi_1,\phi_2,\phi_3)$ remote state preparation cannot be achieved with fidelity $\mathcal{F}$$\geq$$10^{-2}$ via the modified protocol. This in turn points out that complete operational characterization of global unitary operations cannot be obtained in context of remote state preparation.
\section{conclusion}\label{conc}
Role of quantum discord as a measure of quantum correlations is largely debated owing to its failure to remain invariant under arbitrary local measurements\cite{rem6,rem7}. Even then, as already discussed before, it has been studied extensively both from theoretic and practical viewpoints. Our observations have clearly pointed out utility of this feature to reveal optimum effectiveness of global unitary operations in context of extracting nonclassicality of quantum systems which otherwise behave classically. Hence it is now verified that no general(non trivial) two qubit zero discord state can be characterized by the notion of absoluteness. Precisely speaking, given any two qubit zero discord state(other than maximally mixed state), there always exists a suitable global unitary operation(more specifically nonlocal unitary operation $\widehat{\mathcal{U}}$) under which the resulting state has non vanishing quantum discord. Effective nonlocal unitary operations are provided for every possible classical-quantum state(Table.\ref{table:ta3}) and quantum-classical state(Table.\ref{table:ta4}). So, given a two qubit state, one now knows at least one global unitary operation for which a non zero quantum discord can be obtained from the given state. \\
Discord being a useful resource in various quantum information processing tasks, global unitary operations turn out to be effective for increasing utility of classical-quantum or quantum-classical states which otherwise cannot be used in the corresponding tasks. This in turn leads to an operational interpretation of global basis change as discussed in context of remote state preparation task. However there also exist quantum-classical states for which our protocol fails(in sense of achieving fidelity at least of order $10^{-2}$ ). In this context one may recall that such interpretation relies on operational characterization of geometric discord as introduced in \cite{rem1}. So one may try to first operationally characterize quantum discord in perspective of remote state preparation task. Such observations in turn may be further analyzed for characterizing global unitary operations. Besides, studying role of global basis change in perspective of other information processing tasks(involving quantum discord) also provides source of potential future research.\\

\section{Appendix.A}
\textit{Part of proof of Theorem.1}: Here we analyze the effect of nonlocal unitary operations $\widehat{\mathcal{U}}$ on state $\rho_{\mathcal{AB}}^{(1)}$(Eq.(\ref{ist})). This part of the proof is based on the necessary and sufficient condition that $\mathbb{D}_{\rho_{\mathcal{AB}}^{(2)}}(\mathcal{B}/\mathcal{A})$ vanishes if and only if it can be expressed as a classical-quantum state(Eq.(\ref{cq})). As indicated in the main text, every possible form of $\rho_{\mathcal{AB}}^{(1)}$ as a classical-quantum state is given by Table.(\ref{table:ta1}). Now for each of those forms, if possible, let us assume that $\rho_{\mathcal{AB}}^{(2)}$(Eq.(\ref{ist})) can be expressed as a classical-quantum state(Eq.(\ref{cq})). To be precise, we assume existence of unit vector $\vec{u}$$=$$(u_1,u_2,u_3)$ giving direction of projector($\Pi_i^{\mathcal{A}}$) corresponding to
classical part of $\rho_{\mathcal{AB}}^{2}.$ Now under this assumption, $\forall i,k,j,l$$\in$$\{0,1\}$ coefficient of $|ik\rangle\langle jl|$ of $\rho_{\mathcal{AB}}^{(2)},$ $C_{ikjl}$(say) should be equal to that of coefficient of $|ik\rangle\langle jl|$ corresponding to classical-quantum state form of $\rho_{\mathcal{AB}}^{(2)},$ $C_{ikjl}^{\small{CQ}}$(say). Given a $\rho_{\mathcal{AB}}^{(2)},$ failing to obtain equality($C_{ikjl}$$=$$C_{ikjl}^{\small{CQ}}$) for at least one $(i,k,j,l)$ indicates that such a comparison is impossible which in turn proves that our assumption is wrong: $\rho_{\mathcal{AB}}^{(2)}$ is not a classical-quantum state. Consequently $\mathbb{D}_{\rho_{\mathcal{AB}}^{(2)}}(\mathcal{B}/\mathcal{A})$ turns out to be non zero.\\
Now Table.\ref{table:ta1} indicates two possible forms of classical quantum states(Eq.(\ref{cq})).
\begin{equation}\label{cqf1}
\small{\rho_{\mathcal{AB}}^{(1)}}=\small{\frac{1}{4}(\mathbb{I}_{2\times2}+\vec{\mathfrak{m}}.\vec{\sigma}\otimes \mathbb{I}_2+\mathbb{I}_2\otimes \vec{\mathfrak{n}}.\vec{\sigma})}
\end{equation}
 and
\begin{equation}\label{cqf2}
\small{\rho_{\mathcal{AB}}^{(1)}}=\small{\frac{1}{4}(\mathbb{I}_{2\times2}+\mathfrak{m}_i\sigma_i\otimes \mathbb{I}_2+\mathbb{I}_2\otimes \vec{\mathfrak{n}}.\vec{\sigma}+\mathfrak{s}_{ii}\sigma_{i}\otimes\sigma_{i})}
\end{equation}
where $\vec{\mathfrak{m}}$$=$$(\mathfrak{m}_1,\mathfrak{m}_2,\mathfrak{m}_3)$ and $\vec{\mathfrak{n}}$$=$$(\mathfrak{n}_1,\mathfrak{n}_2,\mathfrak{n}_3)$ are real vectors. \\In Eq.(\ref{cqf2}), the index $i$$\in$$\{1,2,3\}.$
Corresponding possible cases are as follow:
\begin{enumerate}
  \item $\vec{\mathfrak{m}}$$=$$(m_1,0,0),$ $\mathcal{S}$$=$$\textmd{diag}(s_{11},0,0)$, $\vec{\mathfrak{n}}$ arbitrary
  \item $\vec{\mathfrak{m}}$$=$$(0,m_2,0),$ $\mathcal{S}$$=$$\textmd{diag}(0,s_{22},0)$, $\vec{\mathfrak{n}}$ arbitrary
  \item $\vec{\mathfrak{m}}$$=$$(0,0,m_3),$ $\mathcal{S}$$=$$\textmd{diag}(0,0,s_{33})$, $\vec{\mathfrak{n}}$ arbitrary
\end{enumerate}
We now start with the first form(Eq.(\ref{cqf1})). For arbitrary $ \vec{\mathfrak{n}},$ depending on $\vec{\mathfrak{m}}$ in Eq.(\ref{cqf1}) following cases are possible: \\
\begin{enumerate}\label{enu1}
  \item $\vec{\mathfrak{m}}$$=$$\vec{\Theta}$
  \item $\vec{\mathfrak{m}}$$=$$(0,0,m_3)$
  \item $\vec{\mathfrak{m}}$$=$$(0,m_2,0)$
  \item $\vec{\mathfrak{m}}$$=$$(m_1,0,0)$
   \item $\vec{\mathfrak{m}}$$=$$(m_1,0,m_3)$
  \item $\vec{\mathfrak{m}}$$=$$(m_1,m_2,0)$
  \item $\vec{\mathfrak{m}}$$=$$(0,m_2,m_3)$
  \item $\vec{\mathfrak{m}}$$=$$(m_1,m_2,m_3)$
\end{enumerate}
Firstly let us consider the trivial subcase of case.1 where both $\vec{\mathfrak{m}}$ and $\vec{\mathfrak{n}}$$=$$\Theta$. This corresponds to the maximally mixed state:$\rho_{\mathcal{AB}}^{(1)}$$=$$\frac{1}{4}\mathbb{I}_{2\times 2}.\,$Clearly after application of any nonlocal unitary operation, $\rho_{\mathcal{AB}}^{(1)}$ remains unchanged. Consequently $\rho_{\mathcal{AB}}^{(2)}$ in this case is a classical quantum state thereby having vanishing discord.\\
We now approach with all possible nontrivial subcases related to each of the above cases starting with that of Case.1.\\
Possible subcases of Case.1 are:\\
\begin{itemize}
  \item $\vec{\mathfrak{n}}$$=$$(n_1,0,0)$
  \item $\vec{\mathfrak{n}}$$=$$(0,n_2,0)$
  \item $\vec{\mathfrak{n}}$$=$$(n_1,n_2,0)$
  \item $\vec{\mathfrak{n}}$$=$$(n_1,n_2,n_3)$ with $n_3$$\neq$$0$ whereas $n_1$ and $n_2$ arbitrary.
\end{itemize}
\textit{$1^{st}$ subcase of case.1:} Let nonlocal unitary operation $\widehat{\mathcal{U}}$$=$$\widehat{\mathcal{U}}(\phi_1,\phi_2,\phi_3)$ characterized by $\phi_1$$=$$\frac{\pi}{2},$ $\phi_2$$=$$\phi_3$$=$$\frac{\pi}{4}$ be applied on $\rho_{\mathcal{AB}}^{(1)}.$ As stated above, let us now consider coefficient of term $|01\rangle\langle 00|$ of $\rho_{\mathcal{AB}}^{(2)}(C_{0100})$ and that of coefficient of term $|01\rangle\langle 00|$ appearing in assumed classical-quantum state form of $\rho_{\mathcal{AB}}^{(2)}(C_{0100}^{\small{CQ}}).$ Equality $C_{0100}$$=C_{0100}^{\small{CQ}}$ demands:
\begin{equation}\label{cn1}
    n_1(1-u_3^2)=0
\end{equation}
As $n_1$$\neq$$0$ and $u_1^2+u_2^2+u_3^2$$=$$1,$ so $u_3$$=$$\pm 1,$ $u_2$$=$$u_1$$=$$0.$ Again $C_{1000}$$=$$C_{1000}^{\small{CQ}}$ demands:
\begin{equation}\label{cn2}
    n_1(-1+u_1u_2+u_2^2)=0
\end{equation}
Using $u_2$$=$$u_1$$=$$0,$ in Eq.(\ref{cn2}) demands $n_1$$=$$0$ leading to a contradiction. Hence $C_{0100}$$=C_{0100}^{\small{CQ}}$ and $C_{1000}$$=C_{1000}^{\small{CQ}}$ do not hold simultaneously. Consequently for this subcase, $\rho_{\mathcal{AB}}^{(2)}$ obtained from classical quantum state $\rho_{\mathcal{AB}}^{(1)}$ after applying nonlocal unitary operation $\widehat{\mathcal{U}}(\frac{\pi}{2},\frac{\pi}{4},\frac{\pi}{4})$ is not a classical-quantum state. So under application of $\widehat{\mathcal{U}}(\frac{\pi}{2},\frac{\pi}{4},\frac{\pi}{4})$ the classical-quantum state $\rho_{\mathcal{AB}}^{(1)}$(Eq.(\ref{cqf1})), characterized by $\vec{\mathfrak{n}}$$=$$(n_1,0,0)$ and $\vec{\mathfrak{m}}$$=$$\Theta,$ gets converted to a `one-way' discord non zero state.\\
\textit{$2^{nd}$ subcase of case.1:} Let nonlocal unitary operation $\widehat{\mathcal{U}}(\frac{\pi}{4},0,\frac{\pi}{2})$ be applied. $C_{0100}$$=$$C_{0100}^{\small{CQ}}$ demands:
\begin{equation}\label{cn3}
  n_2(1-u_3^2)=0
\end{equation}
Hence $C_{0100}$$=$$C_{0100}^{\small{CQ}}$ demands $u_3$$=$$\pm 1,$ $u_2$$=$$u_1$$=$$0.$ But $C_{1101}$$=$$C_{1101}^{\small{CQ}}$ demands:
\begin{equation}\label{cn4}
  n_2(1-u_2^2)=0
\end{equation}
which requires $n_2$$=$$0$ as $u_2$ must be $0$ if $C_{0100}$$=$$C_{0100}^{\small{CQ}}$(Eq.(\ref{cn3})). This again leads to a contradiction as for this subcase, $n_2$$\neq$$0.$ So in this subcase, classical quantum state $\rho_{\mathcal{AB}}^{(1)}$ gets converted to $\rho_{\mathcal{AB}}^{(2)}$ for which $\mathbb{D}_{\rho_{\mathcal{AB}}^{(2)}}(\mathcal{B}/\mathcal{A})$$\neq$$0.$\\
Continuing in this manner, for every possible form of classical-quantum state($\rho_{\mathcal{AB}}^{(1)}$) given by Eq.(\ref{cqf1}) and Eq.(\ref{cqf2}) , it can be shown that applying suitable $\widehat{\mathcal{U}}(\phi_1,\phi_2,\phi_3)$ will generate $\rho_{\mathcal{AB}}^{(2)}$ having $\mathbb{D}_{\rho_{\mathcal{AB}}^{(2)}}(\mathcal{B}/\mathcal{A})$$\neq$$0.$ Instead of putting forward similar arguments, we enlist the suitable required nonlocal unitary operations for all possible subcases of individual cases(as enlisted above) corresponding to first possible form of $\rho_{\mathcal{AB}}^{(1)}$(Eq.(\ref{cqf1})) and also for second possible form given by Eq.(\ref{cqf2}).\\
\section{appendix.B}
Here we discuss the effect of nonlocal unitary operations over all possible forms of quantum-classical states(Table.\ref{table:ta2}). As discussed in Appendix.A, here also we enlist those nonlocal unitaries which are effective in generating states having non vanishing $\mathbb{D}_{\rho_{\mathcal{AB}}^{'}}(\mathcal{A}/\mathcal{B})$ starting from quantum-classical states $\rho_{\mathcal{AB}}.$ As enlisted in Table.\ref{table:ta2}, one of the forms of quantum-classical state(after application of suitable local unitaries) is given by Eq.(\ref{cqf1}) while the other is given by:
\begin{equation}\label{qcf2}
\small{\rho_{\mathcal{AB}}^{(1)}}=\small{\frac{1}{4}(\mathbb{I}_{2\times2}+\vec{\mathfrak{m}}.\vec{\sigma}\otimes\mathbb{I}_2 +\mathbb{I}_2\otimes \mathfrak{n}_i\sigma_i+\mathfrak{s}_{ii}\sigma_{i}\otimes\sigma_{i})(i=1,2,3)}
\end{equation}
With eight possible forms(as listed in Appendix.A) of quantum-classical states corresponding to Eq.(\ref{cqf1}), the possible cases as given by Eq.(\ref{qcf2}) are:
\begin{enumerate}
  \item $\vec{\mathfrak{n}}$$=$$(n_1,0,0),$ $\mathcal{S}$$=$$\textmd{diag}(s_{11},0,0)$, $\vec{\mathfrak{m}}$ arbitrary
  \item $\vec{\mathfrak{n}}$$=$$(0,n_2,0),$ $\mathcal{S}$$=$$\textmd{diag}(0,s_{22},0)$, $\vec{\mathfrak{m}}$ arbitrary
  \item $\vec{\mathfrak{n}}$$=$$(0,0,n_3),$ $\mathcal{S}$$=$$\textmd{diag}(0,0,s_{33})$, $\vec{\mathfrak{m}}$ arbitrary
\end{enumerate}
We now enlist the effective nonlocal unitaries for all possible cases in Table.\ref{table:ta4}.\\
$$\,$$
\begin{center}
\begin{table}
\caption{Details of nonlocal unitary operations to be applied on any possible classical-quantum state having forms given by Eq.(\ref{cqf1}) and Eq.(\ref{cqf2}) so that resulting state has non zero $\mathbb{D}_{\rho_{\mathcal{AB}}^{'}}(\mathcal{B}/\mathcal{A}).$ Observing segregation of all possible cases corresponding to Eq.(\ref{cqf1}) and that to Eq.(\ref{cqf2}), it is easy to interpret that until and otherwise specified, all the components $m_1,m_2,m_3n_1,n_2,n_3$  mentioned in the table are non zero.}
\begin{center}
\begin{tabular}{|c|c|}
\hline
Possible Characterization& $\widehat{\mathcal{U}}(\phi_1,\phi_2,\phi_3)$\\
\hline
 $\vec{\mathfrak{m}}$$=$$\Theta,$ $\vec{\mathfrak{n}}$$=$$(n_1,n_2,n_3)$&$\widehat{\mathcal{U}}(\frac{\pi}{4},\frac{\pi}{4},0)$\\
where $n_3$$\neq$$0$ and $n_1,$$n_2$ arbitrary&\,\\
\hline
$\vec{\mathfrak{m}}$$=$$\Theta,$ $\vec{\mathfrak{n}}$$=$$(n_1,0,0)$&$\widehat{\mathcal{U}}(\frac{\pi}{2},\frac{\pi}{4},\frac{\pi}{4})$\\
\hline
$\vec{\mathfrak{m}}$$=$$\Theta,$ $\vec{\mathfrak{n}}$$=$$(0,n_2,0)$&$\widehat{\mathcal{U}}(\frac{\pi}{4},0,\frac{\pi}{2})$\\
\hline
$\vec{\mathfrak{m}}$$=$$\Theta,$ $\vec{\mathfrak{n}}$$=$$(n_1,n_2,0)$&$\widehat{\mathcal{U}}(\frac{\pi}{4},0,\frac{\pi}{2})$\\
\hline
$\vec{\mathfrak{m}}$$=$$(0,0,m_3),$ $\vec{\mathfrak{n}}$$=$$(n_1,n_2,n_3)$&$\widehat{\mathcal{U}}(0,\frac{\pi}{2},0)$\\
where $n_1$$\neq$$0$ and $n_2,$$n_3$ arbitrary&\,\\
\hline
$\vec{\mathfrak{m}}$$=$$(0,0,m_3),$ $\vec{\mathfrak{n}}$$=$$(0,n_2,n_3)$&$\widehat{\mathcal{U}}(0,\frac{\pi}{2},\frac{\pi}{2})$\\
where $n_2$$\neq$$0$ and $n_3$ arbitrary&\,\\
\hline
$\vec{\mathfrak{m}}$$=$$(0,0,m_3),$ $\vec{\mathfrak{n}}$$=$$(0,0,n_3)$&$\widehat{\mathcal{U}}(0,\frac{\pi}{4},\frac{\pi}{4})$\\
where $n_3$ arbitrary&\,\\
\hline
$\vec{\mathfrak{m}}$$=$$(0,m_2,0),$ $\vec{\mathfrak{n}}$$=$$(n_1,n_2,n_3)$&$\widehat{\mathcal{U}}(0,0,\frac{\pi}{2})$\\
where $n_1$$\neq$$0$ and $n_2,$$n_3$ arbitrary&\,\\
\hline
$\vec{\mathfrak{m}}$$=$$(0,m_2,0),$ $\vec{\mathfrak{n}}$$=$$(0,n_2,n_3)$&$\widehat{\mathcal{U}}(0,\frac{\pi}{2},\frac{\pi}{4})$\\
where $n_2$$\neq$$0$ and $n_3$ arbitrary&\,\\
\hline
$\vec{\mathfrak{m}}$$=$$(0,m_2,0),$ $\vec{\mathfrak{n}}$$=$$(0,0,n_3)$&$\widehat{\mathcal{U}}(\frac{\pi}{4},0,\frac{\pi}{4})$\\
where $n_3$ arbitrary&\,\\
\hline
$\vec{\mathfrak{m}}$$=$$(m_1,0,0),$ $\vec{\mathfrak{n}}$$=$$(n_1,n_2,n_3)$&$\widehat{\mathcal{U}}(0,\frac{\pi}{2},0)$\\
where $n_1$$\neq$$0$ and $n_2,$$n_3$ arbitrary&\,\\
\hline
$\vec{\mathfrak{m}}$$=$$(m_1,0,0),$ $\vec{\mathfrak{n}}$$=$$(0,n_2,n_3)$&$\widehat{\mathcal{U}}(\frac{\pi}{4},\frac{\pi}{2},0)$\\
where $n_2$$\neq$$0$ and $n_3$ arbitrary&\,\\
\hline
$\vec{\mathfrak{m}}$$=$$(m_1,0,0),$ $\vec{\mathfrak{n}}$$=$$(0,0,n_3)$&$\widehat{\mathcal{U}}(0,\frac{\pi}{4},\frac{\pi}{4})$\\
where $n_3$ arbitrary&\,\\
\hline
$\vec{\mathfrak{m}}$$=$$(m_1,0,m_3),$ $\vec{\mathfrak{n}}$ arbitrary&$\widehat{\mathcal{U}}(0,\frac{\pi}{2},\frac{\pi}{2})$\\
\hline
$\vec{\mathfrak{m}}$$=$$(m_1,m_2,0),$ $\vec{\mathfrak{n}}$ arbitrary&$\widehat{\mathcal{U}}(0,\frac{\pi}{2},\frac{\pi}{2})$\\
\hline
$\vec{\mathfrak{m}}$$=$$(0,m_2,m_3),$ $\vec{\mathfrak{n}}$ arbitrary&$\widehat{\mathcal{U}}(\pi,\pi,\frac{\pi}{2})$\\
\hline
$\vec{\mathfrak{m}}$$=$$(m_1,m_2,m_3),$ $\vec{\mathfrak{n}}$ arbitrary&$\widehat{\mathcal{U}}(0,\frac{\pi}{2},\frac{\pi}{2})$\\
\hline
$\vec{\mathfrak{m}}$$=$$(m_1,0,0),$ $\mathcal{S}$$=$$\textmd{diag}(s_{11},0,0)$&$\widehat{\mathcal{U}}(0,0,\frac{\pi}{2})$\\
$\vec{\mathfrak{n}}$ arbitrary&\,\\
\hline
$\vec{\mathfrak{m}}$$=$$(0,m_2,0),$ $\mathcal{S}$$=$$\textmd{diag}(0,s_{22},0)$&$\widehat{\mathcal{U}}(\frac{\pi}{2},\frac{\pi}{2},0)$\\
$\vec{\mathfrak{n}}$ arbitrary&\,\\
\hline
$\vec{\mathfrak{m}}$$=$$(0,0,m_3),$ $\mathcal{S}$$=$$\textmd{diag}(0,0,s_{33})$&$\widehat{\mathcal{U}}(\pi,\frac{\pi}{2},\pi)$\\
$\vec{\mathfrak{n}}$ arbitrary&\,\\
\hline
\end{tabular}
\end{center}
\label{table:ta3}
\end{table}
\end{center}
\begin{center}
\begin{table}
\caption{List of suitable nonlocal unitary operations application of which converts any possible quantum-classical state(forms given by Eq.(\ref{cqf1}) and Eq.(\ref{qcf2})) to $\rho_{\mathcal{AB}}^{'}$ such that $\mathbb{D}_{\rho_{\mathcal{AB}}^{'}}(\mathcal{A}/\mathcal{B}).$ As argued in Table.\ref{table:ta3}, if unspecified, all the components of local Bloch vectors $\vec{\mathfrak{m}}$ and $\vec{\mathfrak{n}}$ mentioned in the table are non zero.}
\begin{center}
\begin{tabular}{|c|c|}
\hline
Possible Characterization& $\widehat{\mathcal{U}}(\phi_1,\phi_2,\phi_3)$\\
\hline
 $\vec{\mathfrak{m}}$$=$$\Theta,$ $\vec{\mathfrak{n}}$$=$$(n_1,n_2,n_3)$&$\widehat{\mathcal{U}}(\frac{\pi}{4},0,\frac{\pi}{2})$\\
where $n_3$$\neq$$0$ and $n_1,$$n_2$ arbitrary&\,\\
\hline
$\vec{\mathfrak{m}}$$=$$\Theta,$ $\vec{\mathfrak{n}}$$=$$(n_1,0,0)$&$\widehat{\mathcal{U}}(\frac{\pi}{4},\frac{\pi}{4},0)$\\
\hline
$\vec{\mathfrak{m}}$$=$$\Theta,$ $\vec{\mathfrak{n}}$$=$$(0,n_2,0)$&$\widehat{\mathcal{U}}(\frac{\pi}{4},0,\frac{\pi}{2})$\\
\hline
$\vec{\mathfrak{m}}$$=$$\Theta,$ $\vec{\mathfrak{n}}$$=$$(n_1,n_2,0)$&$\widehat{\mathcal{U}}(\frac{\pi}{4},\frac{\pi}{4},0)$\\
\hline
$\vec{\mathfrak{m}}$$=$$(0,0,m_3),$ $\vec{\mathfrak{n}}$$=$$(n_1,n_2,n_3)$&$\widehat{\mathcal{U}}(0,\frac{\pi}{2},\frac{\pi}{2})$\\
where $n_2$$\neq$$0$ and $n_1,$$n_3$ arbitrary&\,\\
\hline
$\vec{\mathfrak{m}}$$=$$(0,0,m_3),$ $\vec{\mathfrak{n}}$$=$$(n_1,0,n_3)$&$\widehat{\mathcal{U}}(0,\frac{\pi}{2},\frac{\pi}{2})$\\
where $n_3$$\neq$$0$ and $n_1$ arbitrary&\,\\
\hline
$\vec{\mathfrak{m}}$$=$$(0,0,m_3),$ $\vec{\mathfrak{n}}$$=$$(n_1,0,0)$&$\widehat{\mathcal{U}}(0,\frac{\pi}{4},\frac{\pi}{4})$\\
where $n_1$ arbitrary&\,\\

\hline
$\vec{\mathfrak{m}}$$=$$(0,m_2,0),$ $\vec{\mathfrak{n}}$$=$$(n_1,n_2,n_3)$&$\widehat{\mathcal{U}}(0,0,\frac{\pi}{2})$\\
where $n_1$$\neq$$0$ and $n_2,$$n_3$ arbitrary&\,\\
\hline
$\vec{\mathfrak{m}}$$=$$(0,m_2,0),$ $\vec{\mathfrak{n}}$$=$$(0,n_2,n_3)$&$\widehat{\mathcal{U}}(0,\frac{\pi}{2},\frac{\pi}{4})$\\
where $n_2$$\neq$$0$ and $n_3$ arbitrary&\,\\
\hline
$\vec{\mathfrak{m}}$$=$$(0,m_2,0),$ $\vec{\mathfrak{n}}$$=$$(0,0,n_3)$&$\widehat{\mathcal{U}}(\frac{\pi}{4},\frac{\pi}{4},0)$\\
\hline
$\vec{\mathfrak{m}}$$=$$(0,m_2,0),$ $\vec{\mathfrak{n}}$$=$$(0,0,0)$&$\widehat{\mathcal{U}}(\frac{\pi}{4},0,\frac{\pi}{2})$\\
\hline
$\vec{\mathfrak{m}}$$=$$(m_1,0,0),$ $\vec{\mathfrak{n}}$$=$$(n_1,n_2,n_3)$&$\widehat{\mathcal{U}}(0,0,\frac{\pi}{2})$\\
where $n_1$$\neq$$0$ and $n_2,$$n_3$ arbitrary&\,\\
\hline
$\vec{\mathfrak{m}}$$=$$(m_1,0,0),$ $\vec{\mathfrak{n}}$$=$$(0,n_2,n_3)$&$\widehat{\mathcal{U}}(0,\frac{\pi}{2},\frac{\pi}{4})$\\
where $n_3$$\neq$$0$ and $n_2$ arbitrary&\,\\
\hline
$\vec{\mathfrak{m}}$$=$$(m_1,0,0),$ $\vec{\mathfrak{n}}$$=$$(0,n_2,0)$&$\widehat{\mathcal{U}}(\frac{\pi}{4},\frac{\pi}{2},0)$\\
\hline
$\vec{\mathfrak{m}}$$=$$(m_1,0,0),$ $\vec{\mathfrak{n}}$$=$$(0,0,0)$&$\widehat{\mathcal{U}}(0,\frac{\pi}{4},\frac{\pi}{2})$\\
\hline
$\vec{\mathfrak{m}}$$=$$(m_1,0,m_3),$ $\vec{\mathfrak{n}}$ arbitrary&$\widehat{\mathcal{U}}(0,\frac{\pi}{2},\frac{\pi}{2})$\\
\hline
$\vec{\mathfrak{m}}$$=$$(m_1,m_2,0),$ $\vec{\mathfrak{n}}$ arbitrary&$\widehat{\mathcal{U}}(0,\frac{\pi}{2},\frac{\pi}{2})$\\
\hline
$\vec{\mathfrak{m}}$$=$$(0,m_2,m_3),$ $\vec{\mathfrak{n}}$ arbitrary&$\widehat{\mathcal{U}}(\frac{\pi}{2},0,\frac{\pi}{2})$\\
\hline
$\vec{\mathfrak{m}}$$=$$(m_1,m_2,m_3),$ $\vec{\mathfrak{n}}$ arbitrary&$\widehat{\mathcal{U}}(0,\frac{\pi}{2},\frac{\pi}{2})$\\
\hline

$\vec{\mathfrak{n}}$$=$$(n_1,0,0),$ $\mathcal{S}$$=$$\textmd{diag}(s_{11},0,0)$&$\widehat{\mathcal{U}}(0,0,\frac{\pi}{2})$\\
$\vec{\mathfrak{m}}$ arbitrary&\,\\
\hline
$\vec{\mathfrak{n}}$$=$$(0,n_2,0),$ $\mathcal{S}$$=$$\textmd{diag}(0,s_{22},0)$&$\widehat{\mathcal{U}}(0,0,\frac{\pi}{2})$\\
$\vec{\mathfrak{m}}$ arbitrary&\,\\
\hline
$\vec{\mathfrak{n}}$$=$$(0,0,n_3),$ $\mathcal{S}$$=$$\textmd{diag}(0,0,s_{33})$&$\widehat{\mathcal{U}}(0,\frac{\pi}{2},0)$\\
$\vec{\mathfrak{m}}$ arbitrary&\,\\
\hline
\end{tabular}
\end{center}
\label{table:ta4}
\end{table}
\end{center}

\end{document}